# Value-driven Manufacturing Planning using Cloud-based Evolutionary Optimisation[1]

Shuai Zhao[a], Piotr Dziurzanski[b] and Leandro Soares Indrusiak[c]

Real-Time Systems Group, Department of Computer Science, University of York, United Kingdom

[a]shuai.zhao@york.ac.uk, [b]piotr.dziurzanski@york.ac.uk, [c]leandro.indrusiak@york.ac.uk

**Abstract.** This paper considers manufacturing planning and scheduling of manufacturing orders whose value decreases over time. The value decrease is modelled with a so-called value curve. Two genetic-algorithm-based methods for multi-objective optimisation has been proposed, implemented and deployed to a cloud. The first proposed method allocates and schedules manufacturing of all the ordered elements optimising both the makespan and the total value, whereas the second method selects only the profitable orders for manufacturing. The proposed evolutionary optimisation has been performed for a set of real-world-inspired manufacturing orders. Both the methods yield a similar total value, but the latter method leads to a shorter makespan.

**Introduction**

Manufacturing process planning and scheduling are widely studied subjects [1], but the new reality brought by highly dynamic scenarios proposed under the *Industry 4.0* concept requires a different approach. The common approaches based on simplified models of fairly static industrial configurations, such as the classic Job-shop Scheduling Problem (JSP) and its many extensions, do not match well the type and scale of scenarios found in modern manufacturing plants. Realistic production planning needs to be able to cope with multiple objectives, changing plant conditions, changing production costs and a wide range of potential customer needs.

Meta-heuristics have been used successfully to address more realistic planning and scheduling problems, and among them evolutionary algorithms are particularly well suited to address such problems [2]. Specifically, evolutionary optimisation is the application of an evolutionary algorithm to iteratively uncover improved solutions to an optimisation problem. It is heuristic in nature, meaning that there is no guarantee that it will ever find an optimal solution, or that it will identify a solution as optimal if it is found. Nonetheless, given enough resources it can find a sufficiently fit solution that can be acceptable despite being suboptimal. In this paper, we propose a novel approach to allocate resources to an evolutionary optimization engine for the commodities whose value decreases over time. This approach is applicable to the situations when a price that a customer is willing to pay for the manufactured commodity decreases based on the manufacturing ending time.

Traditionally, search-based manufacturing optimisation has been performed guided by a simulator to evaluate the value of found solutions [3, 4]. Yet such simulation-based optimisers are notorious for long response times when compared to analytical models [5]. In contrast, analytical techniques are usually computed quicker thanks to the application of explicit mathematical formulas and numerical computation methods [1]. Simulation-based evaluators are usually customised for particular use-cases and are difficult to be applied to other scenarios [6]. Despite these drawbacks, analytical techniques are still rather rarely applied to performance evaluation during the manufacturing optimisation. This is very different to other application domains, such as complex computing systems, where analytical methods are applied broadly [7]. Nevertheless, several analytical alternatives has been proposed for manufacturing domain. For example, in [8] a simplistic 3-step fitness function evaluation algorithm has been proposed. Although that paper has not considered such features as

---

multi-modal resource behaviour or multi-objective optimisation, its authors addressed these features as future work.

The algebraic formalism named Interval Algebra has been originally introduced for computer-system resource scheduling in [2]. This algebra has been extended to express the entities and relations found in smart factories in [9]. In this paper, this formalism is also applied, but in contrast to the previous works, the value gained from a task execution changes over time, as specified with a so-called value curve, described in the following section.

**Problem Specification**

In this paper, our objective is to investigate an evolutionary optimisation that is able to cope with the arrival of multiple orders for production processes submitted by potentially different customers. Each one of those production processes is composed of a number of dependent jobs, each of which can be realised in at least one of the machines available in the manufacturing plan. There are different types of machines in a plant and each job can only be executed by a subset of machines in a plant, possibly of various types. Each machine can operate in one of a set of modes, each mode differing in processing time and economic costs. Some machines cannot be used at the same time. Certain sequences of two jobs, scheduled to be processed subsequently by the same machine, can require a time gap of a certain length between them (corresponding to e.g. cleaning the machine in a physical plant).

To capture the dynamic nature of *Industry 4.0* scenarios, we assume that orders do not have a fixed deadline or a fixed cost to the customer. Instead, we assume that each customer can declare how much valuable it is for them if their order is manufactured by a specific point in time. The value $V$ stemming from the manufactured commodities to an end user can therefore be plotted as a value curve $VC(t)$, such as the one depicted in Fig.1. Typically the value is non-increasing, starts with the largest possible value $V_{max}$ when the manufacturing order arrives at time AT. The value remains fixed up to a certain point D, when it starts to change nonincreasingly, for example following a linear trend. The value reaches zero at time point Z, when the delivery is no longer valueable to the customer. Penalties for missing deliveries can also be modelled by curves where the value becomes negative over time. In the figure, the order is finished at time *ET*, so its value to the customer equals $VC(ET)$.

Given a set of orders and the state of the production plant, the goal of the presented evolutionary optimisation is to obtain a production plan and schedule that is able to maximise the overall value obtained by the submitted orders and to minimise the total manufacturing time (makespan).

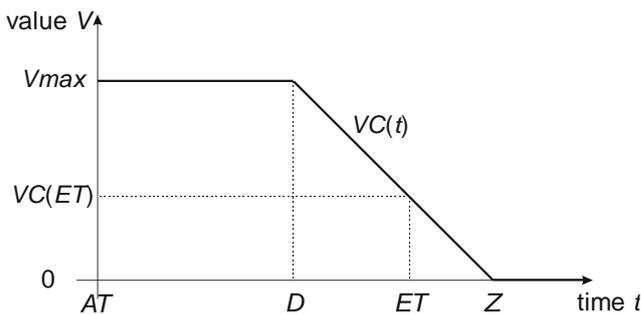

Figure 1. An example value curve of a manufacturing order

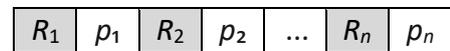

Figure 2. Genes in a chromosome for manufacturing processes

**Proposed Approach**

The key capability of the proposed approach is the ability to respond to dynamic reconfiguration requests. Functionally, the optimiser takes as input an instantaneous description of the manufacturing process and outputs a process plan and schedule. The proposed plan and schedule are high quality, as determined by the fitness function based on Interval Algebra [2]. The job allocation and scheduling is performed by a state-of-the-art multi-objective optimisation genetic algorithm named MOEA/D [10]. In genetic algorithms, candidate solutions are treated as individuals. During the optimisation process, these individuals are evolved in a series of generations, using a set of bio-inspired operations, such as selection, cross-over and mutation. In the considered problem each gene assumes a value from a certain, predefined domain, such as machines identifiers, mode identifier or a priority selected for a particular job. Hence, the so-called value encoding of chromosomes can be applied. There is one to one correspondence between a job and the target machine and mode. The role of the optimiser is to allocate the jobs to resources & modes and schedule them in time. The encoding has hence to embrace both the spatial and temporal scheduling. Consequently, in the proposed encoding a chromosome contains genes of two types, as shown in Fig. 2. For *n* jobs that need to be scheduled, the number of genes is thus equal to 2*n*. The odd *n* genes indicate the target resource (*R*) and its mode, whereas the remaining *n* genes specify the priority (*p*), which influences the ordering of the manufacturing process: the ready jobs of with a higher priority are manufactured earlier. Such chromosome is then forwarded to a fitness function, where the value gained from each job is evaluated using Interval Algebra [2]. The optimisation ends after reaching a predefined number of generations.

Two versions of the optimiser are proposed. In the first one, named *MO Standard*, all ordered elements must be manufactured, regardless of whether manufacturing of a certain element is profitable (i.e., leads to a non-zero value) or not. In the second versions, named *MO with selection*, only the profitable elements are manufactured. The evaluation and comparison of these two versions are presented in the following section.

The optimiser is available as a Docker container. Hence it can be deployed in a cluster with the Kubernetes container-orchestration system, which is available in all major cloud facilities, including AWS, Azure, CloudStack, GCE, OpenStack, OVirt, Photon, VSphere, IBM Cloud Kubernetes Service, as well as can be installed locally. It results in not only the full flexibility in selecting the data centre for deployment, but also benefits from numerous cloud-computing features, such as load balancing or autoscaling. It means that the number of optimiser containers can be dynamically changed to answer the current users' requirements. This feature is possible due to the fact that the optimiser containers are designed to be stateless, i.e., they do not store any session-related data that shall be persistent. Such a distributed, container-based architecture of the proposed solution is in line with the state-of-the-art software design and deployment.

**Experiments**

In this section, a real-world manufacturing scenario is used that demonstrates job allocation and scheduling via a value-driven multi-objective optimisation process. The manufacturing of each ordered element can be produced by a set of machines with various working modes. Table 1 gives an example of available manufacturing ways for one ordered element, where each machine with a working model yield to a unique execution time and a max profit for production of the given element. In general, for each machine, a longer execution time of an ordered part indicates higher manufacturing quality, and hence, comes with a higher max profit. However, note that the actual profit for each ordered part also depends on the finish time, as given by Figure 1.

Table 1. An example configuration of an ordered element

| Element | Machine Number | Working Mode | Execution Time (s) | Max Profit ($) |
|---|---|---|---|---|
| E1 | M1 | Mode 1 | 2833.5 | 167.0 |
| | | Mode 2 | 2956.2 | 168.4 |
| | | Mode 3 | 3042.1 | 175.9 |
| | | Mode 4 | 3174.1 | 192.1 |
| | M2 | Mode 1 | 2033.5 | 230.0 |
| | | Mode 2 | 2156.2 | 237.1 |
| | | Mode 3 | 2242.1 | 238.6 |
| | | Mode 4 | 2674.1 | 273.1 |
| | M3 | Mode 1 | 1256.2 | 481.6 |
| | | Mode 2 | 1633.5 | 462.1 |
| | | Mode 3 | 1842.1 | 519.3 |
| | | Mode 4 | 1974.1 | 596.9 |

The optimisation has been performed for two objectives: makespan (lower is better) and total profit (higher is better). In the experiment scenario, 14 assorted elements have been ordered for manufacturing. The population size has been set to 300 individuals and the number of generations has been limited to 500. Unless specified otherwise, the value curve is configured as D=30,000s and Z=40,000s for each ordered part. we later on apply different value curve configurations and investigate its impact on the optimisation results. An example Pareto front approximation obtained after 500 generations is presented in Fig. 3. From the obtained non-dominated solutions, we observed that the two considered objectives (i.e., misnaming makespan and maximising total profit) explicitly contradict each other, where a higher makespan often yielded to a higher total profit.

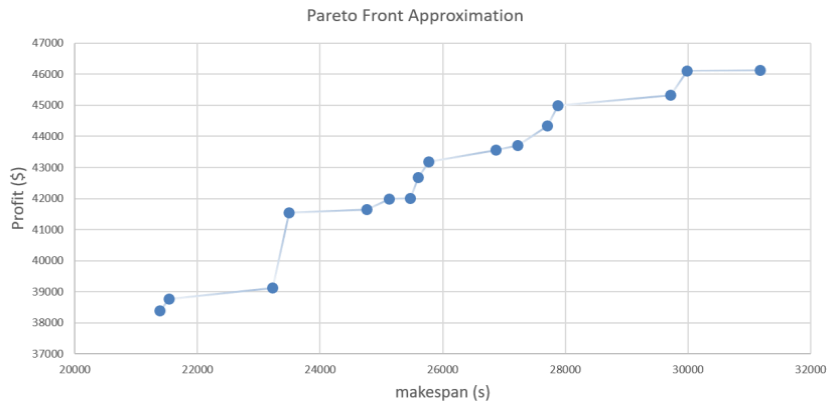

Figure 3. Pareto front of an example 14-element order optimisation

In the next experiment, 200 example orders for manufacturing 7 to 16 elements (10 orders for each) have been analysed. Fig. 4 presents the optimisation results for both the objectives for these orders for the MO Standard approach. As shown in this figure, both the makespan and the total profit follow an increasing trend, but the total profit grows slower after $i \geq 10$. The reason for it is the actual profit of each ordered part depends heavily on the manufacturing ending time. As some elements have to wait for their manufacturing, their actual profit can be much lower than the maximal profit, or even equal to 0. This observation may lead to the conclusion that manufacturing fewer numbers of elements can be similarly beneficial, but the makespan can be lower.

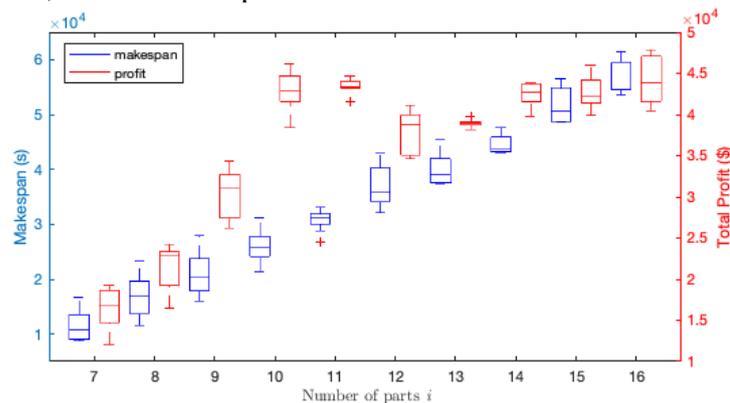

Figure 4. Makespan and the total profit of analysed 200 example orders for the MO standard method

The above hypothesis has been evaluated experimentally and the results are presented in Figure 5 for all considered 200 example orders. In general, MO with selection has led to a slightly higher total profit than MO standard in all cases. However, MO with selection yields much lower make span than that of the MO standard can achieve, especially with $i \geq 12$, for rejecting ordered parts that are less likely to be profitable due to the delayed manufacturing time. To further demonstrate the effectiveness of the MO with selection approach, various value curve configurations have be applied to the example orders of assorted sizes, with optimisation results shown in Table 2. In general, MO with selection has led to slightly higher total profits than MO standard (by up to 15.9%, with D=30,000s and

Z=40,000s), but its makespan has been, in average, 50.1% shorter than the one obtained with MO standard. This significant difference has been caused by the rejection of the 28.6% of elements to be manufactured as being unprofitable. Similar relations between these two methods have been observed for all the scenarios.

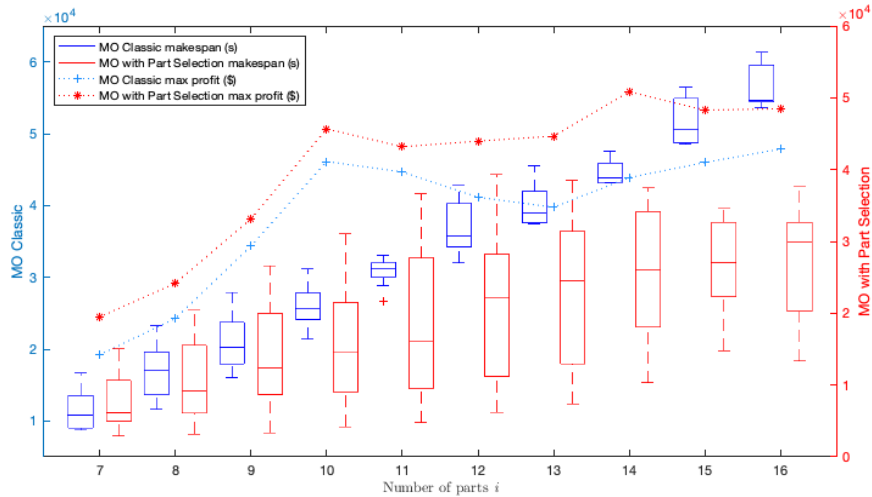

Figure 5. Makespan of 200 example orders for the two analysed optimisation methods

Table 2. Comparison between MO Standard and MO with selection for various value curve shapes

| D | Z | MO with selection | | | MO standard | | |
|---|---|---|---|---|---|---|---|
| | | Profit ($) | Make-span (s) | Elements produced | Profit ($) | Make-span (s) | Elements produced |
| 5000 | 10000 | 12174 | 9299 | 6 | 12965 | 51787 | 14 |
| 5000 | 15000 | 16321 | 10690 | 7 | 16293 | 53555 | 14 |
| 10000 | 15000 | 19286 | 12402 | 8 | 18842 | 46537 | 14 |
| 10000 | 20000 | 22262 | 16195 | 9 | 21391 | 49345 | 14 |
| 15000 | 20000 | 23787 | 18292 | 9 | 23816 | 46537 | 14 |
| 15000 | 25000 | 27777 | 24028 | 10 | 26131 | 47293 | 14 |
| 20000 | 25000 | 31137 | 23434 | 9 | 31446 | 48692 | 14 |
| 20000 | 30000 | 35243 | 26125 | 10 | 34807 | 48692 | 14 |
| 25000 | 30000 | 38005 | 25556 | 10 | 36616 | 51894 | 14 |
| 25000 | 35000 | 41880 | 32090 | 12 | 40791 | 51894 | 14 |
| 30000 | 35000 | 43198 | 32090 | 12 | 42781 | 55558 | 14 |
| 30000 | 40000 | 50849 | 37440 | 13 | 43892 | 47562 | 14 |
| 35000 | 40000 | 50849 | 37440 | 13 | 45799 | 47562 | 14 |
| 35000 | 45000 | 53037 | 41397 | 12 | 47032 | 47562 | 14 |

From the experiments it may be concluded that both the proposed optimisation techniques are similar in terms of the obtained total profit, but considering the total makespan, MO with selection is significantly better.

**Conclusions**

In this paper, two Genetic-Algorithm-based methods have been presented for multi-objective optimisation of integrated process planning and scheduling of manufacturing orders whose profit decreases over time. The first of the proposed methods allocates and schedules all the ordered elements so that the makespan and the total profit are optimised. The second method selects only these orders whose manufacturing is assessed as being profitable. Both approaches have been experimentally evaluated. The quality of both methods in terms of the obtained total profit is similar, but the latter method produces fewer ordered elements and hence the plans generated by this method have shorter makespan. The second method can be then concluded to dominate the first one and, as such, should be used for real-world cases similar to the one described in the paper. In future, we plan to extend our approaches to deal with any weakly decreasing value curves and even non-monotonic value curves.

**Acknowledgment**

The authors acknowledge the support of the EU H2020 SAFIRE project (Ref. 723634).

**References**

[1] R. Briesemeister and A. Novaes: Comparing an approximate queuing approach with simulation for the solution of a cross-docking problem. J. of Applied Mathematics, vol. 2017 (2017), p. 11.

[2] L. S. Indrusiak and P. Dziurzanski: An interval algebra for multi-processor resource allocation. Int. Conf. on Embedded Computer Systems: Architectures, Modeling, and Simulation (SAMOS) (2015), pp. 165–172.

[3] D. Ivanov: Operations and supply chain simulation with AnyLogic: Decision-oriented introductory notes for master students. Berlin School of Economics and Law (preprint) (2017).

[4] R.Liu, X.Xie, K.Yu, and Q.Hu: A survey on simulation optimization for the manufacturing system operation. International Journal of Modelling and Simulation, no. 2, vol. 38 (2018), pp. 116–127.

[5] H. Zisgen, I. Meents, B. R. Wheeler, and T. Hanschke: A queueing network based system to model capacity and cycle time for semiconductor fabrication. Winter Simulation Conference, (2008), pp. 2067–2074.

[6] G. Shao, A. Brodsky, and R. Miller: Modeling and optimization of manufacturing process performance using Modelica graphical representation and process analytics formalism. Journal of Intelligent Manufacturing, no. 6, vol. 29 (2018), pp. 1287–1301.

[7] A.E. Kiasari, A. Jantsch, and Z. Lu: Mathematical formalisms for performance evaluation of networks-on-chip. ACM Computing Survey, no. 3, vol. 45 (2013), pp. 38:1–38:41.

[8] V. Nguyen and H. Bao: An efficient solution to the mixed shop scheduling problem using a modified genetic algorithm. Procedia Computer Science, vol. 95 (2016), pp. 475–482.

[9] P. Dziurzanski, J. Swan, L.S. Indrusiak, J.M. Ramos: Implementing Digital Twins of Smart Factories with Interval Algebra. IEEE Int. Conf. on Industrial Technology (2019).

[10] Q. Zhang and H. Li: MOEA/D: A Multiobjective Evolutionary Algorithm Based on Decomposition. IEEE Trans. on Evolutionary Computation, no. 6, vol. 11 (2007), pp. 712-731.

[11] S. Zhao, P. Dziurzanski, L.S. Indrusiak: Value-driven Manufacturing Planning using Cloud-based Evolutionary Optimisation. International Conference on Manufacturing Science and Technology, ICMST 2019.